\documentclass[12pt,twoside]{article}
\usepackage{fleqn,espcrc1}
\usepackage{graphicx}
\usepackage[figuresright]{rotating}

\newcommand{\AmS}{{\protect\the\textfont2
  A\kern-.1667em\lower.5ex\hbox{M}\kern-.125emS}}
\title{Chiral symmetry breaking in the Wegner-Houghton RG approach}

\author{A. Bonanno \address{Osservatorio Astrofisico di Catania, Via S. Sofia
          58, I-95125, Catania\\
INFN, Sezione di Catania, Corso Italia 57, I- 95129, Catania}
        and
	D. Zappal\`a 
        \address{INFN, Sezione di Catania and
	Dip.to di Fisica e Astronomia, Universit\`a di Catania,\\        
        Corso Italia 57, I-95129, Catania, Italy}
       \thanks{Talk given by D. Zappal\`a}}
\begin{document}

\maketitle

\begin{abstract}
The Wegner-Houghton formulation of the exact renormalization group
evolution equation is used in order to study the chiral symmetry breaking
of the linear $\sigma$-model coupled to an isospin doublet of quarks.
A  numerical investigation for a particular truncation of 
the equation which includes the scalar field renormalization function 
is presented. 
\end{abstract}

\section{INTRODUCTION}

 The full understanding of the phase structure  of QCD as the theory of 
strong interactions has become an important issue since the 
discovery of asymptotic freedom. The possibility that high 
temperature QCD could show different properties from the theory at zero 
temperature has been addressed already in \cite{coll}. New features are also 
predicted for the theory at very high and intermediate baryon density 
\cite{bail}, \cite{alfo,rapp}. 

An essential role is played by the Chiral Symmetry
which is supposed to be broken by the vacuum structure of the theory.
Thus, before considering the high temperature phase transition 
with the Chiral Symmetry restoration, one has to deal with the 
problem of determining a nonzero order parameter which indicates the Chiral 
Symmetry Breaking (CSB) at zero temperature and density. A strong 
simplification is obtained by considering the 
effective theory described by a linear $\sigma$ model of scalar mesons 
coupled to an isospin multiplet of fermions with the same chiral 
symmetry group of the original action. 
Then CSB corresponds to a mexican hat-shaped 
effective potential and the order parameter is identified with the vacuum 
expectation value (VEV) of one of the scalar fields.

The infrared (IR) properties of a field theory
such as the vacuum structure, can be analysed by means of the wilsonian 
renormalization group which generates a sequence of effective actions 
defined at some momentum scale $k$, starting from the  original action, 
by integrating out in the latter all the modes with frequency 
higher than $k$. This leads to the construction of a differential flow 
equation for the $k$-dependent action, generally known as Exact 
Renormalization Group Equation (ERGE). 
In the past years
many analytical and numerical approaches to the ERGE, applied to various
theories, have been developed. Two recent extensive reviews of this topic are 
\cite{rep1,rep2} and a detailed ERGE analysis of the Chiral Phase Transition 
can be found in \cite{berg2,jung,berg3,rep3}.

Here we shall consider just the simple case of the CSB at zero temperature
employing the sharp cut-off version of the ERGE, namely the Wegner-Hougton
equation \cite{wegn}. The cut-off dependence of the ERGE for  the O(4) linear
$\sigma$-model has been addressed in\cite{papp} where, however
only the flow  of the scalar potential is considered while the  Yukawa 
coupling is kept fixed. In addition to the Yukawa coupling flow, 
we are particularly interested in the behavior of the  
field renormalization. In fact, if our model is really an 
effective theory which, at some scale, can naturally replace the original 
QCD action, then at this scale one expects the formation of the mesonic 
bound states characterized by a very small wave function renormalization
function.
The reduction of QCD to the  quark-meson action has been 
analysed before and a review of it is given in \cite{rep3}. 
As expected the value of the scalar field renormalization 
is vanishingly small at the mesonic bound state formation 
scale which is found between 0.60 and 0.63 GeV. 
Then, the flow of the field renormalization deserves particular attention. 

\section{ANALYSIS AND RESULTS}

The explicit form of the euclidean action at a scale $k$ is  
($\rho\equiv\sigma^2+\vec\pi^2$)
\begin{equation}
S_k=\int d^4x \Biggl( {{z_l}_k\over 2}
\partial_\mu \sigma\partial_\mu \sigma+
{{z_t}_k\over 2}
\partial_\mu \vec \pi\partial_\mu \vec\pi+
U_k(\rho)
+\sum_c^{N_c} \bar\psi_c \bigl (i\gamma\cdot \partial+g_k\sigma
+i g_k \gamma^5 \vec\pi \vec\tau \bigr )\psi_c \Biggr)
\label{eq:leq}
\end{equation}
$\psi_c$ is a  flavor doublet and the summation over the index $c$ is 
extended to $N_c$ colors. No CSB fermion mass term is 
included and the action is fully invariant under chiral transformations. 
The presence of  CSB depends on the shape of the O(4) symmetric potential 
$U_k$ in the limit $k\rightarrow 0$. ${z_l}_k$ and ${z_t}_k$ are the $\sigma$
and $\pi$ renormalization functions. As long as $U_k$ has zero vacuum 
expectation value (VEV), we choose the 
following parametrization $U_k=(1/2)m_k^2\rho+(\lambda_k/24)\rho^2$ and 
consider
the flow of the two parameters $m^2_k$ and $\lambda_k$ neglecting all the 
irrelevant operators that could be generated during the flow. In the presence 
of a nonvanishing VEV $\bar\rho_k$, according to \cite{papp}, 
we reparametrize the potential in terms of $\bar\rho_k$  and $\lambda_k$, 
recalling that $m^2_k$ can be expressed in term of these two parameters 
through the minimum condition.
The purpose of this double choice is practical, we take the couple 
of parameters which corresponds to the simplest set of flow equations.

The Wegner-Houghton equation can be reduced to a set of 
coupled ordinary differential 
equations for the $k$ dependent parameters in (\ref{eq:leq}),
i.e. $g_k$, ${z_l}_k$, ${z_t}_k$, $\lambda_k$ and $m^2_k$ or 
$\bar\rho_k$. In particular the equations for the parameters in the 
potential are deduced in \cite{papp}, and the one for $g_k$ can be obtained 
by an analogous procedure. The equations for ${z_l}_k$, ${z_t}_k$, (which 
coincide as long as $\bar\rho_k=0$, according to the chiral simmetry)
are deduced following the procedure formulated in \cite{fras} and 
already implemented in \cite{noiz}. 
In the following, the derivatives of  ${z_l}_k$ and ${z_t}_k$
w.r.t. the fields, which are supposed to be small, are neglected.

The pion decay constant  and the constituent fermion mass are two IR 
constraints for our equations. In fact, in the limit $k\rightarrow 0$ we have 
$\sqrt{{z_l}_k {\bar\rho}_k} \equiv f \rightarrow f_\pi \sim 0.09$ Gev
and $g_k \sqrt{{\bar\rho}_k}\equiv M \rightarrow M_q \sim 0.3$ GeV,
as initial conditions for the equations.
If we start, according to \cite{rep3}, the flow at the UV scale $k=0.6$ GeV
with a convex potential ($m^2_k>0$), then at a certain scale 
$K_\chi$, $m^2_k$ becomes negative generating a nonzero $\bar\rho_k$.
In the following the three remaining initial conditions imposed are 
$\bar\rho_k=0$ at the CSB scale $K_\chi$  and a particular value for ${z_l}_k$
at the UV scale (the symmetry requires ${z_l}_k={z_t}_k$  at the UV scale).
\begin{figure}[thb]
\begin{minipage}[t]{76mm}
\flushleft\includegraphics[width=79mm]{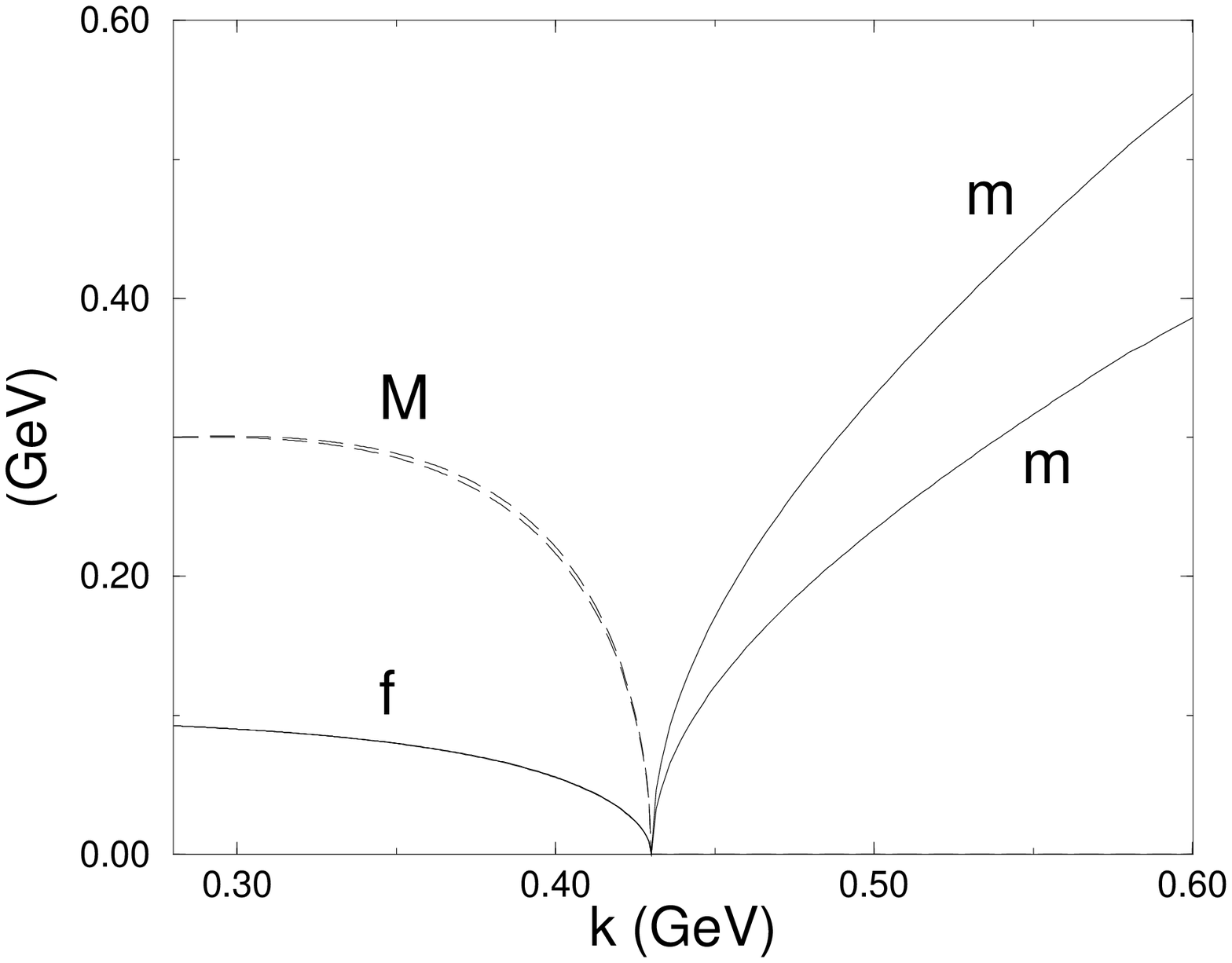}
\caption{RG flow with two different sets of initial conditions (see text). 
For these two sets, the quark mass $M$ and the decay constant $f$ are 
displayed below $K_\chi=0.43 GeV$, and the scalar mass $m$ above $K_\chi$.}
\label{fig:uno}
\end{minipage}
\hspace{\fill}
\begin{minipage}[t]{76mm}
\flushleft\includegraphics[width=79mm]{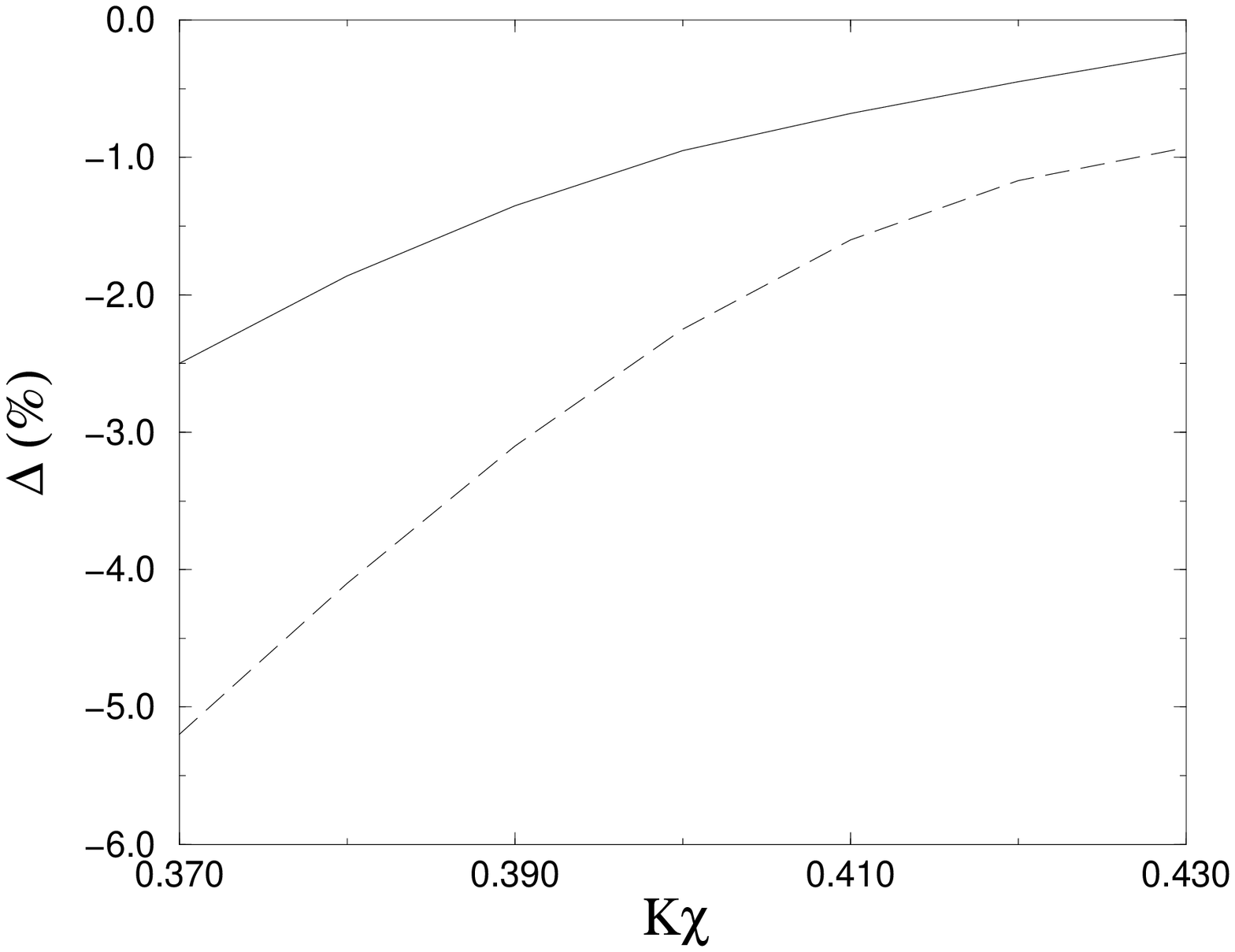}
\caption{$100\times\Delta$ (see text)  vs. $K_\chi$ (GeV) for two different 
values of $M_q$. $M_q=0.30~GeV$ (solid) and $M_q=0.33~GeV$ (dashed).}
\label{fig:due}
\end{minipage}
\end{figure}

Figure \ref{fig:uno} shows, above $K_\chi$ (fixed at the value $K_\chi=0.43$ 
GeV), the flow of the scalar mass $m$ (the subscript $k$ is omitted)
for two different initial values of $z_l(=z_t)$ and, below $K_\chi$,
the corresponding $f$ and $M$. The curves for $f$ and $M$ in the two cases 
are almost superposed indicating insensitivity to the UV condition on $z_l$.
The value of $K_\chi$ is suggested by an IR stability criterion. 
In order to use a particular value of quark mass, $M_q$, as 
the initial condition for the running quantity $M$ defined above, $M$ must be 
practically flat around $k=M_q$. The same criterion should apply to 
$f_\pi$ but, since $f_\pi$ is smaller than $M_q$, 
we find that $f$ at the scale $k=f_\pi$ is already a stable parameter. After 
fixing the IR condition $M(k=M_q-\epsilon)=M_q$ with $\epsilon=0.03$ GeV, 
we have considered the ratio $\Delta=(M(k=M_q)-M_q)/M_q$, plotted vs. $K_\chi$
in Figure \ref{fig:due} for two different values of $M_q$. 
The stability requirement indicates that $K_\chi$ should not be smaller 
than $\sim 0.4 $ Gev. 

\begin{figure}[htb]
\begin{minipage}[t]{76mm}
\flushright\includegraphics[width=79mm]{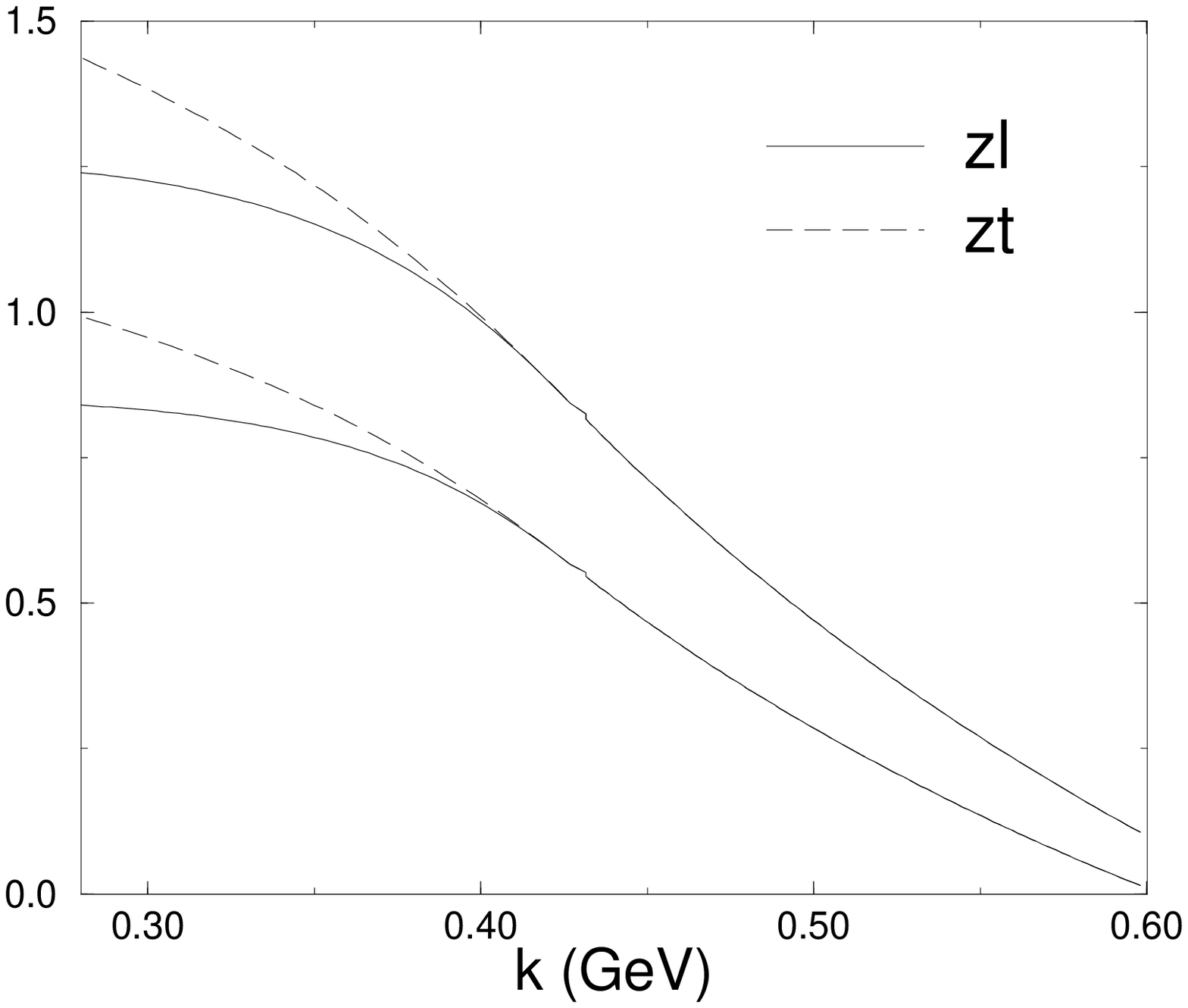}
\caption{Flow of $z_l$ (solid) and $z_t$ (dashed) for two
different sets of initial data (see text). $z_l$ and $z_t$
coincide above $K_\chi=0.43$.}
\label{fig:tre}
\end{minipage}
\hspace{\fill}
\begin{minipage}[t]{76mm}
\flushright\includegraphics[width=79mm]{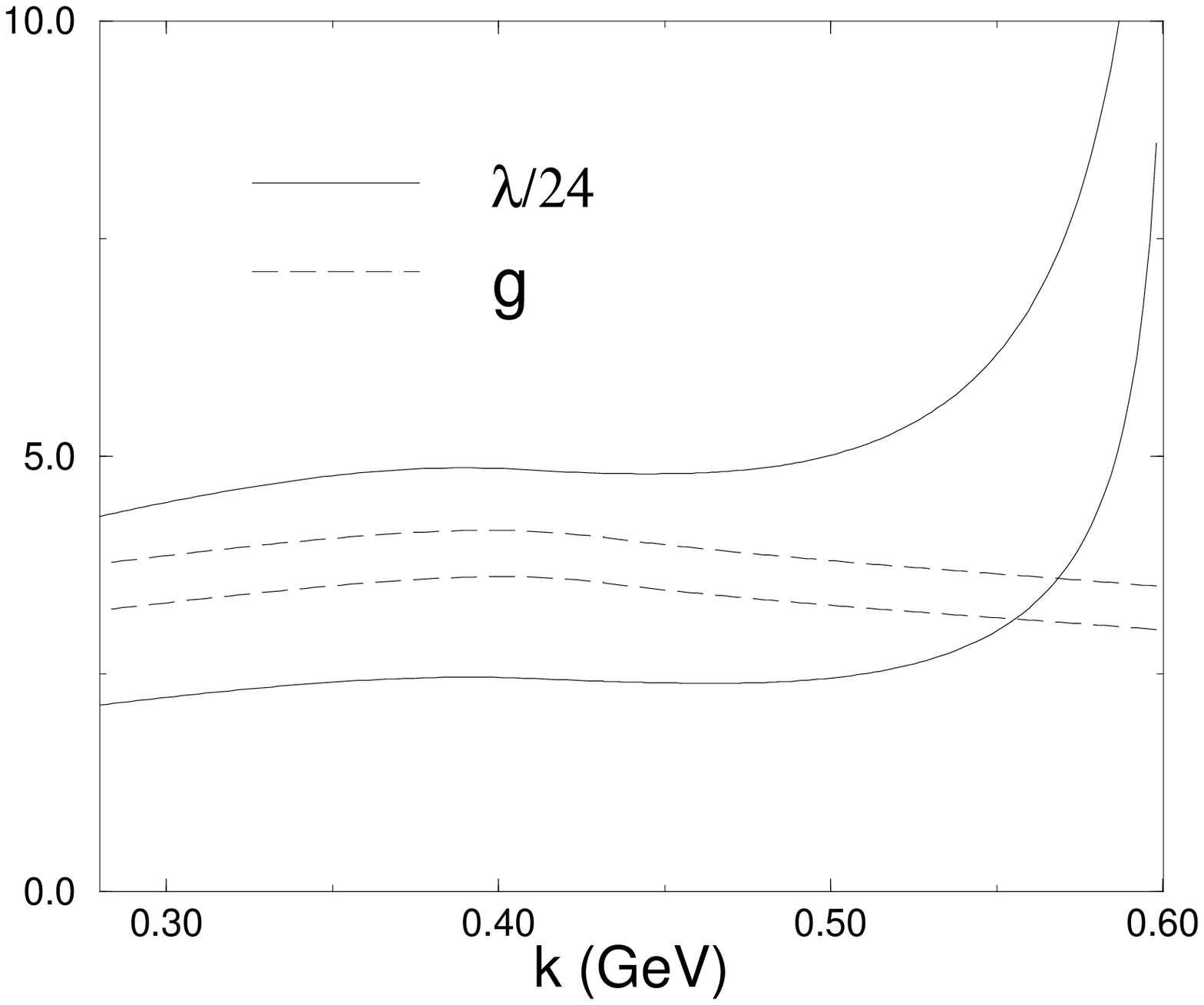}
\caption{Flow of the scaled scalar quartic coupling $\lambda/24$ (solid) and 
of the Yukawa coupling $g$ (dashed) for two different sets of initial data
(see text).}
\label{fig:quattro}
\end{minipage}
\end{figure}

In Figures \ref{fig:tre} and \ref{fig:quattro} the other running parameters
are displayed for the same two initial values of $z_l$  that have been used 
in Figure \ref{fig:uno}. It should be noted the small running of $g$ and 
$\lambda$, in the  range of $k$ considered, and their large values which are 
clearly nonperturbative.
As shown in Figure \ref{fig:quattro}, when going from the UV to the IR 
scale, the field renormalizations grow and, below $K_\chi$,
$z_l$ and $z_t$ increase differently. Since reasonably we expect, for the 
pion physics at scales  around $k=0.3$ GeV, field renormalization
functions close to the unity, we must require very small values at 
$k=0.6$ GeV. Namely we used $z_l=z_t=0.1$ and $z_l=z_t=0.01$ in the 
two cases here considered in Figures \ref{fig:uno},\ref{fig:tre},
\ref{fig:quattro}. Higher curves for $m, \lambda,g$ correspond to 
the former case, lower curves to the latter.

The approximation here employed for the ERGE has been sufficient for 
exploring a nonperturbative region (note that, as explained in \cite{rep3},
the renormalized couplings which include the field renormalization effect,
are larger than those in Figure \ref{fig:tre}) and it has yielded a very small
(vanishing) scalar field renormalization around $0.6$ GeV.
We hope to extend our analysis to the finite temperature case.

\end{document}